\documentclass[twocolumn,showpacs,preprintnumbers,amsmath,amssymb,prl]{revtex4}

\usepackage[dvips]{graphicx}
\usepackage{dcolumn}
\usepackage{amssymb}
\usepackage{amsfonts}
\usepackage{bm}

\begin{document}

\preprint{APS}

\title{Elegance of disordered granular packings:  a validation of Edwards' hypothesis}

\author{Philip T. Metzger}
\email{Philip.T.Metzger@nasa.gov}
\affiliation{%
The KSC Applied Physics Laboratory, John F. Kennedy Space Center, NASA\\
YA-C3-E, Kennedy Space Center, Florida  32899
}%

\author{Carly M. Donahue}
\affiliation{%
Department of Physics, Astronomy and Geology, Berry College\\
2277 Martha Berry Hwy. NW, Mount Berry, GA  30149
}%

\date{\today}

\begin{abstract}
We have found a way to analyze Edwards' density of states for static granular packings in the special case of round, rigid, frictionless grains assuming constant coordination number.  It obtains the most entropic density of single grain states, which predicts several observables including the distribution of contact forces.  We compare these results against empirical data obtained in dynamic simulations of granular packings.  The agreement is quite good, helping validate the use of statistical mechanics methods in granular physics.  The differences between theory and empirics are mainly related to the coordination number, and when the empirical data are sorted by that number we obtain several insights that suggest an underlying elegance in the density of states.
\end{abstract}

\maketitle

The intriguing behaviors of sand and other granular materials are not well understood from a fundamental point of view \cite{Halsey} and there is no theory with a pedigree equivalent to the Navier-Stokes or Maxwell's equations to explain how their state evolves over time, even in the most commonly occurring scenarios.  Considering the ubiquity of granular materials in nature, this is quite surprising.  Making a new effort to explain their physics, Edwards and Oakeshott proposed that the methods of statistical mechanics may be successfully applied \cite{edwards1}.  They hypothesized \textit{a priori} a flat measure in the statistical ensemble---that every metastable arrangement of grains (a \textit{blocked} state) is equally probable under common conditions---and that the analysis of this ensemble should predict some of the important behaviors.  

Because dynamics of granular materials are nonlinear, lossy, and quite different than the dynamics of atoms, it is an important question whether they are ergodic or whether they bias the measure such that Edwards' hypothesis would not be correct.  Seeking to answer this, a number of empirical tests have been performed by computer simulation.  In these, Edwards' hypothesis has successfully predicted packing behaviors for several idealized models \cite{tapping} and the diffusion-mobility behavior of individual grains when a simulated packing is slowly sheared \cite{makse}.  In both cases it appears that the dynamics cause the geometry of the model to explore some region of the phase space with sufficient ergodicity to justify the flat measure.  Also, experiments vibrating a powder have shown that it achieves a steady-state volume, repeatable but dependant on the frequency and amplitude of the vibration \cite{chicago}.  Edwards has used that as the starting point to develop a Boltzmann equation \cite{edwardsentropy}, which assumes the individual grains occupy volumes of space that are statistically uncorrelated to that of their neighbors.  Surely for friction-dominated packings (such as powders) this is reasonable, and so Edwards' transport equation proves ergodicity in the Boltzmannian sense.

\begin{figure}
\includegraphics[angle=0,width=0.45\textwidth]{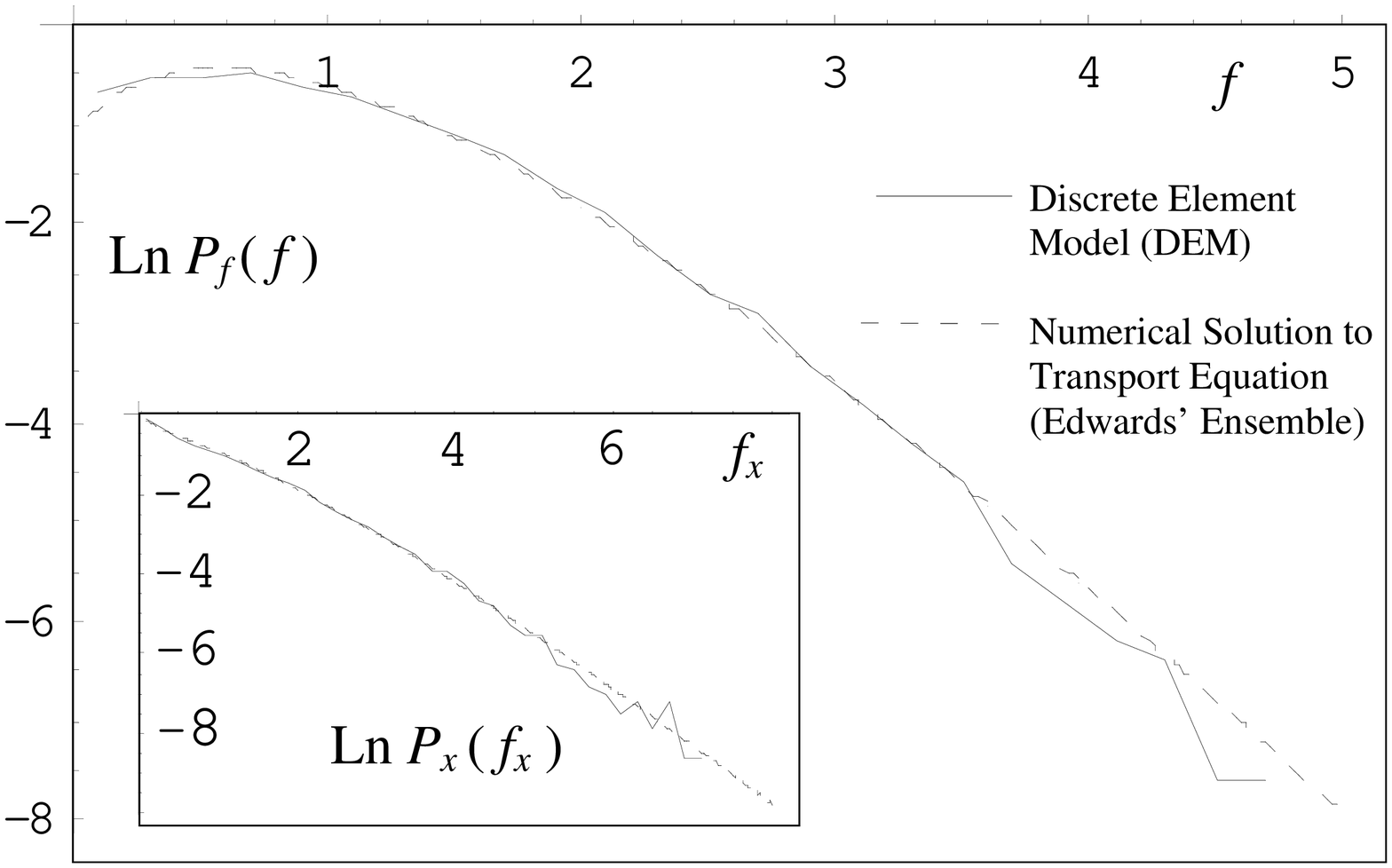}
\caption{\label{fig1} (Semilogarithmic) Distribution of granular contact force magnitudes $P_{f}\left(f\right)$ (main graph) and Cartesian components $P_x(f_x)$ (inset).  The theory and empirical discrete element model (DEM) are strikingly in agreement.  This implies that Edwards' hypothesis is sufficient to capture important organizational features of quasi-static granular physics.}
\end{figure}
In this Letter we present a different kind of test for Edwards' hypothesis.  Rather than examining the geometric features of the packing, we demonstrate that Edwards' flat measure correctly predicts the distributions for single grain load states and for contact forces. This prediction is centrally important to a statistical mechanics theory because the distribution reflects how the ensemble is organized and demonstrates whether or not the correct physics have been incorporated.  In particular, Edwards' hypothesis should predict at least three features in the contact force distribution $P_f(f)$ as illustrated in Fig.~\ref{fig1}:  the wide tail \cite{metzger1} related to the heterogeneity of stresses in a packing (\textit{force chains}) \cite{forcechains}; the small peak near the average value of force under isotropic conditions \cite{antony} related to static equilibrium of the grains (\textit{jamming}) \cite{jamming}; and the non-zero probability density at zero force, $P_f(0)>0$, related to the tipping of grains (\textit{fragility}) \cite{fragile}.  The combination of these three features is unique to the granular distribution, not being found in the typical densities of states for thermal systems.  If Edwards' hypothesis fails to produce this form then it is doubtful that it could become the basis for a theory of quasi-static rheology, since the tipping or sliding of individual grains depends upon the state of their contact forces.

Following Edwards and coworkers \cite{edwardsstress}, we focus on the case of amorphous packings of cohesionless, rigid grains all having the same coordination number $Z$ that makes the packing isostatic \cite{isostatic}.  Our case is further idealized by using two dimensional round grains with monodisperse grain diameters, omitting gravity and working in the thermodynamic limit (infinitely large packings) so that the boundary layer may be neglected.  We focus on the frictionless case so that $Z=4$, and we limit this Letter to isotropic stress and fabric although our methodology can solve for anisotropic cases, too.  The idealizations may be taken out in future refinements of the theory, but this is a good starting point because packings of cohesionless, round grains that are perfectly rigid \cite{radjai} and/or frictionless \cite{frictionless, silbertgrest} and/or monodisperse \cite{antony, silbertgrest} have been the focus of many empirical studies and are known to have force distributions with the same features as the less idealized packings.  Hence they must be subject to the same basic organization in the physics.

The goal of the analysis is to combine Edwards' microcanonical DOS \cite{edwardsgeo} and contact force probability functional \cite{edwardsstress} and then derive the density of single grain states $\rho_{\text{g}}(w_x,w_y,\theta_1,\ldots,\theta_4)$.  The first two arguments of this density, the \textit{Cartesian loads}, are
\begin{equation}
w_{x}=\frac{1}{2}\sum_{\gamma=1}^{4} f_{\gamma} |\cos \theta_{\gamma}|,\ \ \ w_{y}=\frac{1}{2}\sum_{\gamma=1}^{4} f_{\gamma} |\sin \theta_{\gamma}|.
\end{equation}
where $f_\gamma$ and $\theta_\gamma$ are the four contact force magnitudes and contact angles on a grain.  In the special case we have selected, solving for $\rho_{\text{g}}$ provides everything that can be known about the individual grains in the packing.  For example, it contains the joint distribution of contact forces and angles,
\begin{eqnarray}
P_{f\theta}(f,\theta) & = & \int_{0}^{\infty}\!\!\!\text{d}^{2}w \int_{0}^{2\pi}\!\!\!\text{d}^{4}\theta\ \rho_{\text{g}}\times \frac{1}{4}\sum_{\gamma=1}^{4}\delta(\theta-\theta_{\gamma}) \nonumber\\*
& & 
\times\ \delta\left[f-f_{\gamma}(w_x,w_y,\theta_1,\ldots,\theta_4)\right] \label{collapse2},
\end{eqnarray}
and the fabric distribution $P_{4\theta}(\theta_1,\ldots,\theta_4)$ \cite{troadec}.

The analytical method \cite{metzger2} is to count states in Edwards' ensemble and maximize entropy applying the same well-known procedure that has been used to derive the Bose or Fermi distributions \cite{textbook}.  The result is
\begin{eqnarray}
\rho_{\text{g}}(w_{x},w_{y},\theta_1, \ldots, \theta_4) = G(\theta_1,\ldots,\theta_4)\ e^{-\lambda_{x}w_{x}-\lambda_{y}w_{y}} \nonumber\\*
\times \prod_{\gamma=1}^4 [P_{f\theta}\left(f_{\gamma},\theta_{\gamma}\right)]^{1/2}\ \Theta\left(f_{\gamma}\right),\ \ \label{eqnp2w4th}
\end{eqnarray}
where $\Theta$ is the Heaviside (unit step) function, $\lambda_x$ and $\lambda_y$ are the Lagrange multipliers that scale mechanical loading, and $G$ derives from the array of Lagrange multipliers used to conserve $P_{4\theta}$.  Eqs.~(\ref{collapse2}) and~(\ref{eqnp2w4th}) form a recursion in $P_{f\theta}$ and $\rho_{\text{g}}$, the ``transport'' equation, which may be solved numerically using $P_{4\theta}$ and the mechanical loading as inputs.  

For the present the transport equation has been solved in the isotropic case with a simplifying approximation:  
\begin{eqnarray}
\rho_g(w_x,w_y,\theta_\beta) \approx \rho_w(w_x,w_y)\rho_\theta(\theta_\beta)\nonumber\\*
\times \Theta_S(w_x,w_y,\theta_\beta),
\end{eqnarray}
where $\Theta_S$ is a function that evaluates either to unity or zero if the grain is stable or unstable, respectively \cite{metzger3}.  This modified separability assumes no correlation between the loads and fabric apart from the truncating effect of $\Theta_S$.  The physical idea is that correlation does arise predominantly because nature disallows unstable grains.  Empirical results have shown this to be correct \cite{metzger2}.  In the remainder of this Letter, ``the theory'' refers to the resulting numerical solution.

\begin{figure}
\includegraphics[angle=0,width=0.45\textwidth]{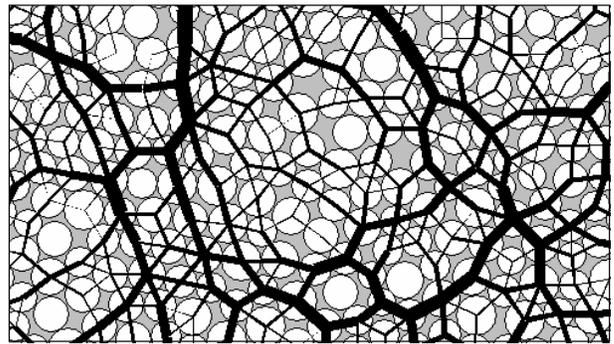}
\caption{\label{packing} Portion of a Discrete Element Model (DEM) showing the disordered packing fabric and propagation of force chains.  Line width is proportional to force magnitude.  Although disordered, a simple pattern can be discerned in the density of states as discussed in the text.}
\end{figure}

To compare with the theory, we have performed discrete element modeling (DEM) \cite{DEM} of 17,000 two dimensional, round, frictionless grains.  A portion of our DEM packing is shown in Fig.~\ref{packing} to contrast the spatial disorder of its force network with the simplicity of the statistics, discussed below.
The grain diameters are uniformly distributed from $1.0$ to $1.5$ to reduce crystallization.  The grains were deposited isotropically into a square test cell with hard walls and without gravity, and their diameters increased by rescaling, producing the desired isotropic stress state.  The grains were allowed to move dynamically until they located and settled into a blocked state.  They have a linear spring contact law, but staying near the jamming transition avoids excessive deformation of the contacts and approximates the grain rigidity of the analysis.  Data from grains in the boundary layer (chosen to be 4 grain diameters along each wall) were discarded to reduce the boundary effects, which we found will otherwise significantly skew the statistics.  We also note that the theory assumes $Z=4$ for every grain, but the DEM produces significant populations for $Z=3$, 4, and 5.  Therefore, to evaluate Edwards' hypothesis we will discuss the differences between these populations.

Fig.~\ref{fig1} shows the DEM data compared to the theory for $P_f(f)$ and for the distribution of Cartesian components of force, $P_x(f_x)$.  They are in remarkable agreement, demonstrating all the correct features and thereby indicating that the ensemble naturally incorporates the correct contact force physics.

\begin{figure}
\includegraphics[angle=0,width=0.45\textwidth]{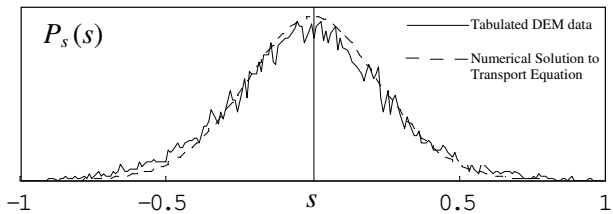}
\caption{\label{sdist} Distribution of $s=(w_x-w_y)/t$ related to the shear stress borne by single grain states.  Solid curve---DEM.  Dashed curve---theoretical prediction, the numerical solution of the transport equation.}
\end{figure}

\begin{figure}
\includegraphics[angle=0,width=0.45\textwidth]{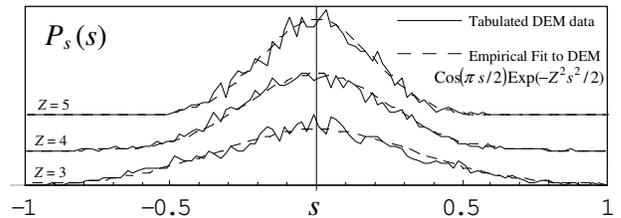}
\caption{\label{scompares} The same distribution $s$ from the DEM as in Fig.~\ref{sdist} except segregated by grain coordination number $Z$ into three graphs (shifted vertically for clarity).  An empirical fit was suggested by the theory (dashed curves) which fits all three $Z$ populations when the standard deviation $\sigma=1/Z$ is the only parameter.}
\end{figure}

To investigate the DOS more fully we note that $w_x$ and $w_y$ are not statistically independent and therefore we would need to plot their statistics as a joint distribution.  However, the change of variables to $t=w_x+w_y$ and $s=(w_x-w_y)/t$ achieves (approximate) statistical independence so that they can be separated more meaningfully.  The parameter $t$ is analogous to hydrostatic pressure but at the grain scale whereas $s$ is a ratio that indicates the degree of shear stress at the grain scale.  The distribution of the latter, $P_s(s)$, in Fig.~\ref{sdist},
demonstrates remarkable agreement between the DEM data and the theory.  We can fit them to a functional form, $P_s(s) = \cos(\pi s/2) \exp(-s^2/2\sigma^2)$ with $\sigma=1/4$.  To explore the dependence on $Z$ we segregated the DEM data into $Z=3$, 4, and 5 populations and plotted $P_s(s)$ for each in Fig.~\ref{scompares}.  
Remarkably, a good fit to each population is made simply by writing $\sigma=1/Z$.  This identifies a previously unknown pattern in the form of the DOS.  

This result has an interesting relationship with recent work on the statistics of cooperative bridges.  These bridges naturally occur within the bulk of packings and have been a focus of much interest due to the way that they direct the propagation of stress.  In their recent work, Mehta \textit{et al.} found that the lengths of these bridges have an exponential distribution like $P_f(f)$, and so bridges were proposed to be a geometrical analog of the force chains themselves \cite{Mehta}.  At the same time, they found the spatial orientations of the bridges to have a Gaussian distribution, which is similar to $P_s(s)$.  It has been pointed out to us that this bridge orientation is in fact closely related to $s$ because the angle with respect to the gravity vector determines the shear stress borne by a bridge.  Thus, it appears that both the exponential and the Gaussian statistics found within the single grain DOS may be connected, through the mesoscopic feature of bridges, to important macroscopic behaviors.

The other statistically independent variable, $t$, was also analyzed in the theory and found to have the distribution
\begin{equation}
P_t(t)=t^{\beta-1} e^{-\beta t}\label{PofT}
\end{equation}
where $t$ has been normalized and where $\beta = 5$.  This is an extremely interesting form when several facts are considered.  First, it is well known in probability theory that, when several independent random variables $t_i$ having distributions $P_i$ are added together, $T=\sum_i t_i$, then the distribution of the sum is 
\begin{equation}
P_T=P_1 \otimes P_2 \otimes \cdots \otimes P_n
\end{equation}
where $\otimes$ is the convolution operator.  Second, it is well known that Eq.~\ref{PofT} is the convolution of pure exponentials,
\begin{equation}
t^{\beta-1} e^{-\beta t}=e^{-t} \otimes e^{-t} \otimes \cdots \otimes e^{-t}, t>0
\end{equation}
where there are $\beta$ exponentials being convolved, to be precise.  Third, a pure exponential is of course the canonical (Gibbs) distribution.  Together, these facts tell us that the hydrostatic load on a grain in a disordered packing is distributed as if it were composed of several statistically independent, canonical contributions.  This is quite surprising because the contact forces themselves are neither independent nor canonical.

\begin{figure}
\includegraphics[angle=0,width=0.45\textwidth]{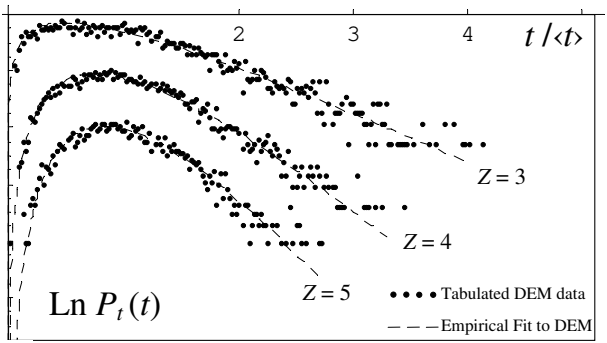}
\caption{\label{tdist2} Distribution of $t$ (hydrostatic loading of the grains) from the DEM, segregated into three graphs by their coordination number $Z$ (shifted vertically for clarity).  The empirical fits are $P_t(t)=t^{\beta-1} e^{-\beta t}$ as predicted by the theory, using $\beta=2Z-4$.}
\end{figure}

To check this, we segregated the DEM data by $Z$ and obtained $P_t(t)$ for each population as shown in Fig.~\ref{tdist2}.
This confirmed the pattern:  all three populations are fit perfectly by Eq.~\ref{PofT}, using $\beta=2Z-4$ as the only parameter.  While the origin of the value of $\beta$ is yet to be explained, it is clear that the essential physics have been correctly incorporated into this theory because the forms of the distributions are all correctly predicted.  

Furthermore, an important feature of $\beta$ can be seen:  averaging over all the grains in the packing, $\langle \beta \rangle = \langle Z \rangle$ if and only if $\langle Z \rangle = 4$.  This also happens to be the condition for mechanical isostaticity and recent studies have demonstrated that it really is satisfied for the present case \cite{isostatic}.  This means that, if the independent canonical variables suggested by Eq.~\ref{PofT} can be identified, we will find that the number of them is exactly equal to the number of contact forces in the packing.  This is surprising because in general $\beta \ne Z$ and therefore the new variables cannot be localized to the individual contacts.  This forms an interesting analogy to the molecular vibrations in a solid, which are resolvable into non-localized phonon statistics, or to the eigenmodes of a mass-spring network.

In summary, the theory predicts the correct forms for $P_f(f)$, $P_x(f_x)$, $P_s(s)$ and $P_t(t)$.  By segregating the grains of a dynamic simulation by their coordination number $Z$, we discover that all the populations fit the theory's predicted DOS (represented by $s$ and $t$) with only a simple parameter change based on $Z$.  This identifies a previously unrecognized but elegant pattern.  

We conclude that all of the features of $P_f(f)$ (as produced by dynamic simulations and experiments) are naturally predicted by Edwards' hypothesis, alone; none of these features are the result of dynamically-induced departures from a flat measure.  Therefore, Edwards' hypothesis, without recourse to the individual grains' dynamics, produces an ensemble that contains the force chains, the fragility, and all the other important granular phenomena that have been correlated to those features of $P_f(f)$.  The results of this Letter therefore provide one more indication that Edwards' hypothesis may be the correct starting point for a complete statistical mechanics theory of granular packings.

\begin{acknowledgments}
We are grateful for helpful discussions with Aniket Bhattacharya of the University of Central Florida Physics Department and with Robert Youngquist of NASA's John F. Kennedy Space Center.
\end{acknowledgments}


\begin{thebibliography}{99}
\bibitem{metzger3}  Note that unstable grains are cases where $f_\gamma(w_x,w_y,\theta_1,\ldots,\theta_4)<0$ for any $\gamma=1,\ldots,4$.
\end{thebibliography}
\end{document}